\documentclass[twocolumn,showpacs,prb,preprintnumbers,amsmath,amssymb]{revtex4}

\usepackage{dcolumn}
\usepackage{bm}

\usepackage{graphicx}     
\begin{document}

\title{Spontaneous Pseudospin Spiral Order 
in the Bilayer Quantum Hall Systems}
\author{K. Park}
\affiliation{Condensed Matter Theory Center, 
Department of Physics, University of Maryland,
College Park, MD 20742-4111}
\date{\today}

\begin{abstract}
Using exact diagonalization of bilayer quantum Hall systems
at total filling factor $\nu_T=1$ in the torus geometry,
we show that there is a new long-range interlayer phase coherence 
due to spontaneous pseudospin spiral order
at interlayer distances larger than the critical value
at which the pseudospin ferromagnetic order is destroyed.
We emphasize the distinction between
the interlayer phase coherence and
the pseudospin ferromagnetic order.
\end{abstract}

\maketitle

\section{Introduction}
\label{intro}

Bilayer quantum Hall systems
at total filling factor $\nu_T=1$ exhibit 
one of the most interesting many-body correlation effects:
spontaneous interlayer phase coherence,
which is solely caused by the Coulomb interaction
in the limit of zero interlayer tunneling.
Recently this phenomenon has received
drastic experimental support 
by Spielman {\it et al.} \cite{Spielman}, 
who discovered a strong enhancement 
in the zero-bias interlayer tunneling conductance 
for interlayer distance
$d < d_c \simeq 1.83 l_0$ in the regime of very little tunneling, 
where $l_0 = \sqrt{\hbar c/eB}$ is the magnetic length. 
These experiments strongly indicated 
a presence of spontaneous interlayer coherence
at small interlayer distances.

The ground state in the limit of zero interlayer distance
can be shown to be spontaneously interlayer phase coherent
due to layer symmetry.
One of the most convenient way to see this is
to use the pseudospin representation
in which the top (bottom) layer is denoted as 
pseudospin $\uparrow$ ($\downarrow$).
In the limit of zero interlayer separation,
there is exact pseudospin SU(2) symmetry 
so that our bilayer system at $\nu_T =1$ with pseudospin 
can be exactly mapped onto a single layer system $\nu =1$ with real spin.
Then, it is straightforward to see that
the ground state should be a pseudospin ferromagnet
in order to reduce the Coulomb energy cost, i.e. Hund's rule.
On the other hand, it has been known that, 
in the limit of large interlayer separation,
the ground state is composed of two split 
composite fermion Fermi seas \cite{Jain,Kalmeyer,HLR,Scarola}.

Therefore, it is clear that there should be 
a quantum phase transition at a critical interlayer distance $d_c$
since the ground states at small and large interlayer separations
are not adiabatically connected.
However, it is not at all clear what happens near $d_c$.
In spite of intense theoretical efforts in the past,
the true nature of phase transition near $d_c$ has been elusive.
Analytical approaches based on Hatree-Fock theories \cite{Fertig,MacDonald}
were not able to reliably treat strong quantum fluctuations 
near the phase transition,
which led to the erroneous prediction of 
broken translational symmetry. 
On the other hand, 
analytic approaches utilizing effective field theories \cite{Moon,Yang} 
were not directly based on 
the microscopic Hamiltonian of the many-body Coulomb interaction.
So its validity should be justified by other means
such as exact diagonalization.
In particular, mapping of the original many-body Coulomb Hamiltonian 
to a spin model Hamiltonian 
is not expected to be valid 
near the critical interlayer separation.     
Previous numerical studies based on exact diagonalization 
\cite{He,Schliemann}, however, did not take into account 
fundamental fluctuations of interlayer number difference 
due to intrinsic, quantum-mechanical uncertainty in layer indices.
Consequently those studies could not provide
any direct information 
related to spontaneous interlayer phase coherence.

It is crucial at this point to distinguish between 
pseudospin ferromagnetic order and interlayer phase coherence. 
There is no fundamental reason to believe that they are identical
because the interlayer phase coherence is a necessary condition, 
but not a sufficient condition for pseudospin ferromagnetism.
We will show in this article that, at intermediate interlayer distances,
there is another sort of long-range interlayer phase coherence
due to pseudospin spiral ordering.
It should be emphasized that this pseudospin spiral order is
spontaneous in that it exists solely because of the Coulomb interaction
without any external parallel magnetic field.
The formation of pseudospin spiral order
has been considered previously
in the presence of parallel magnetic field \cite{Yang,Demler}.

This article is organized as follows.
We will begin by defining the Hamiltonian of bilayer quantum Hall systems
in Sec. \ref{sec_hamil}. 
The pseudospin magnetization is computed
as a function of interlayer distance $d$
in the limit of zero interlayer tunneling
in Sec. \ref{sec_Sx}
where the intrinsic, quantum-mechanical
fluctuation of interlayer number difference
is fully taken into account,
based on the analogy with superconductivity.
Defined as the point 
where the pseudospin ferromagnetic order is destroyed,
the critical interlayer distance $d_c$ is reliably determined.
In Sec. \ref{sec_spiral} we focus on 
the nature of the instability which destroys
pseudospin ferromagnetic order at $d_c$.
We will show that the instability is due to 
the formation of pseudospin spiral order
which is manifested in the lowest energy excitations.
By computing a new order parameter for pseudospin spiral,
we will show in Sec. \ref{op}
that the ground state itself has 
the pseudospin spiral order. 
Finally, it is argued in Section \ref{conclusions}
that a signature of the phase transition 
from pseudospin ferromagnetic order to our predicted pseudospin spiral order 
may have been already observed experimentally.

\section{Hamiltonian}
\label{sec_hamil}

Let us begin by writing
the Hamiltonian for the bilayer quantum Hall systems:
\begin{equation}
H = H_{t} + \hat{P}_{LLL}V_{Coul}\hat{P}_{LLL}
\label{Hamiltonian}
\end{equation}
where $\hat{P}_{LLL}$ is the lowest Landau level (LLL) projection operator.
$V_{Coul}$ represents the usual Coulomb interaction between electrons:
\begin{equation}
\frac{V_{Coul}}{e^2/\epsilon l_0} = 
\sum_{i<j \in \uparrow} \frac{1}{r_{ij}} 
+\sum_{k<l \in \downarrow} \frac{1}{r_{kl}} 
+\sum_{i \in \uparrow, k \in \downarrow} \frac{1}{\sqrt{r^2_{ik} 
+(d/l_0)^2}}
\end{equation}
where 
$l_0=\sqrt{\hbar c/eB}$ is the magnetic length,
$d$ is the interlayer distance,
and $r_{ij}$ is the lateral separation 
between the {\it i}-th and {\it j}-th electrons.
In the above we have used a pseudospin representation 
to distinguish the top ($\uparrow$) and 
the bottom ($\downarrow$) layers.
In general, we define the pseudospin operator as follows:
\begin{equation}
{\bf S} \equiv 
\frac{1}{2} \sum_{m} c^{\dagger}_{ma} \vec{\sigma}_{ab} c_{mb},
\end{equation}
where $\sigma$ is the usual Pauli matrix, and
$m$ denotes the LLL orbital index. 
Note that $S_z$ is half the electron number difference between the two layers,
and $S_x$ is associated with interlayer tunneling.
We take the real spin to be fully polarized.

The tunneling Hamiltonian $H_{t}$ in Eq.(\ref{Hamiltonian}) 
can be written as:
\begin{equation}
H_{t} = -\frac{t}{2} \sum_{m} \left( c^{\dagger}_{m\uparrow} c_{m\downarrow}
+ c^{\dagger}_{m\downarrow} c_{m\uparrow} \right)
= -t S_x ,  
\label{H_t}
\end{equation}
where $t$ is the single particle interlayer tunneling gap. 
Although Eq.(\ref{H_t}) is valid for general $t$,
we are interested only in the limit of zero interlayer tunneling,
i.e. $t/(e^2/\epsilon l_0) \rightarrow 0$, which is
appropriate when considering spontaneous interlayer coherence
(note that the $t \rightarrow 0$ limit is not the same as the 
$t = 0$ situation). 
We analyze the Hamiltonian in Eq.(\ref{Hamiltonian})
by using exact diagonalization (via a modified Lanczos method)
in the torus geometry \cite{torus}.

\section{Spontaneous pseudospin magnetization}
\label{sec_Sx}

As mentioned in the beginning,
it will be shown that there is a new long-range interlayer phase coherence
due to pseudospin spiral order at sufficiently large $d$. 
But, first, we should determine 
the critical interlayer distance, $d_c$, 
at which pseudospin ferromagnetic order terminates.
Of course, the most natural order parameter for the pseudospin
ferromagnetic order is 
the pseudospin magnetization in the $x$-direction: $\langle S_x \rangle$.

It is tempting to apply exact diagonalization techniques
to finite systems with fixed number of electrons in each layer.
But, $S_x$ changes the interlayer number difference by unity 
($\Delta S_z = \pm 1$) so that 
the ground state expectation value is precisely zero
when computed naively without 
any explicit interlayer tunneling.
This problem has been addressed in an {\it ad hoc} manner 
by introducing explicit interlayer tunneling \cite{Schliemann},
which, however, severely obscures the effect of 
spontaneous phase coherence
and may produce misleading results.
The real solution is to realize that there is
intrinsic, quantum-mechanical uncertainty 
in the layer index of electrons at small interlayer distance 
(even in the limit of zero tunneling)
so that the true ground state $|\psi\rangle$ 
is a linear combination of various states with different $S_z$:
\begin{equation}
|\psi\rangle = \sum_{M} \lambda_M |\phi_M\rangle, 
\end{equation}
where $|\phi_M\rangle$ is the lowest energy state with $S_z=M$, 
$\lambda_M$ is a sharply peaked function of $M$ 
with width ${\cal O}(\sqrt{N})$, and
$N$ is the number of electrons. 
It is important to remember that 
even the Halperin (1,1,1) state, $\psi_{(1,1,1)}$, has
$\lambda_M \propto \exp{(-2 M^2/N)}$,
which can be proved by analyzing the following, explicit wavefunction
including both the orbital and pseudospin part \cite{111}: 
\begin{equation}
|\psi_{(1,1,1)}\rangle =
\prod_m \left[ \frac{1}{\sqrt{2}}(c^{\dagger}_{m\uparrow}
+c^{\dagger}_{m\downarrow}) \right] |0\rangle .
\label{111}
\end{equation}

$\psi_{(1,1,1)}$ is the exact ground state at $d=0$. 
Because it can be exactly mapped onto the BCS wavefunction,
it is fruitful to consider an analogy with the BCS theory
for general $d$.
Following the BCS theory,
one can compute the pseudospin magnetization $S_x$ as follows:
\begin{eqnarray}
\langle\psi|S_x|\psi\rangle &=& 
\sum_{M,M'} \lambda_M \lambda_{M'} 
\langle\phi_{M'}|S_x|\phi_M\rangle 
\nonumber 
\\ 
&\simeq& \langle\phi_{M^*+1}| S_x |\phi_{M^*}\rangle 
+\langle\phi_{M^*-1}| S_x |\phi_{M^*}\rangle 
\nonumber 
\\
&\simeq& 2 \langle\phi_{M^*+1}| S_x |\phi_{M^*}\rangle
\label{BCS}
\end{eqnarray}
where the last two step are justified in the thermodynamic limit
since $\lambda_M$ is sharply peaked at $M=M^*$. 
Although $\langle S_x \rangle$ should be exactly $N/2$ at $d=0$ 
in the thermodynamic limit 
due to the exact pseudospin SU(2) symmety,
there are some finite-system size corrections. 
After some algebra on the (1,1,1) state, 
Eq.(\ref{BCS}) shows that 
$\langle \psi_{(1,1,1)}| S_x | \psi_{(1,1,1)} \rangle$ 
is $(N+1)/2$ for N odd ($M^*=-1/2$), 
and is $\sqrt{N(N+2)}/2$ for N even ($M^*=0$).
Therefore, the pseudospin magnetization 
should be scaled as $N+1$ ($\sqrt{N(N+2)}$) for $N$ odd (even).

\begin{figure}
\includegraphics[width=2.3in]{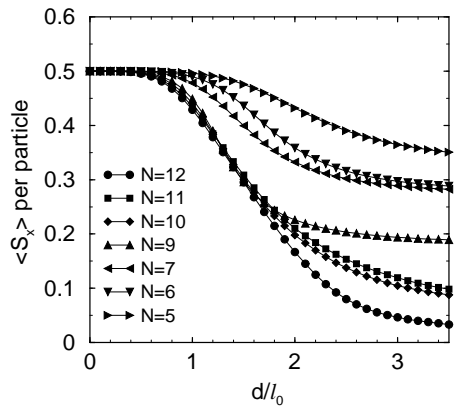}
\caption{Pseudospin magnetization ($S_x$)
as a function of interlayer distance $d/l_0$
for various numbers of electrons $N$.
\label{fig1}}
\end{figure}

Fig.\ref{fig1} shows 
the pseudospin magnetization per particle 
(defined in Eq.(\ref{BCS}))
as a function of $d/l_0$ 
for various numbers of electrons.
As conventional in finite system analysis,
the critical interlayer distance 
can be obtained from
the inflection point of $\langle S_x \rangle$: 
$d_c \simeq 1.5 l_0$.
Although $d_c$ was estimated to be around $1.5 l_0$
previously in various approaches,
it should be emphasized that
the pseudospin magnetization $\langle S_x \rangle$ is computed
for the first time in this article 
without any interlayer tunneling. 
And therefore $d_c$ is reliably determined without any ambiguity.

\section{Pseudospin spiral instability}
\label{sec_spiral}

\begin{figure}
\includegraphics[width=2.5in]{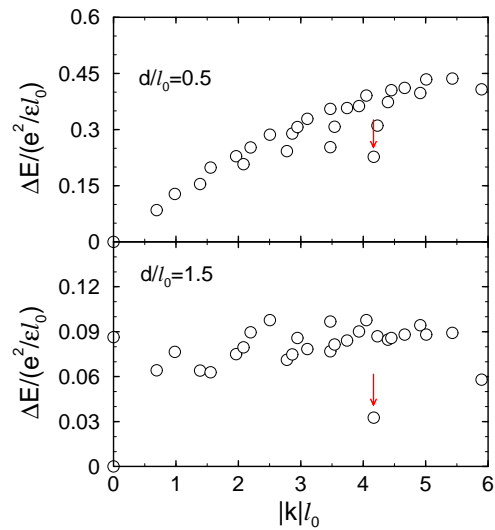}
\caption{Dispersion curves of the lowest energy excitations
at interlayer distance $d/l_0 = 0.5$ and $1.5$
for a finite system with the total number of electrons $N=13$ 
and the number of pseudospin-up electrons $N_{\uparrow}=7$ 
(or $N_{\downarrow}=6$).
It will be shown that
the lowest energy state 
in the momentum channel ${\bf k}l_0=\sqrt{2\pi/N}(N_{\downarrow},0)$
(denoted by arrows)
is the pseudospin spiral state.
\label{fig2}}
\end{figure}

We now address the nature of instability causing 
the destruction of pseudospin ferromagnetic order.  
This question is, in turn, 
directly related to the nature of
the new ground state at $d > d_c$.
To be specific, we consider the lowest energy excitations
which are responsible for the ground state instability.
Fig.\ref{fig2} shows the dispersion curves of the lowest energy excitations
at $d/l_0 = 0.5$ and $1.5$ 
for a finite system with the total number of electrons $N=13$ 
and the number of pseudospin-up electrons 
$N_{\uparrow}=7$ (or $N_{\downarrow}=6$).
There are several points to be emphasized.

First,
there is clearly
a linearly dispersing Goldstone mode
at small interlayer distances, such as $d/l_0 = 0.5$.
However, 
$within numerical accuracy 
$regarding the discreteness of finite-system wavevectors, 
the Goldstone mode seems to vanish for $d \gtrsim d_c$,
which coincides with the destruction of 
pseudospin ferromagnetic order \cite{Nonlocal}.
Consistent with the conclusion from $\langle S_x \rangle$,
this shows that the ground state undergoes a phase
transition around $d=d_c$.

Second, 
contrary to the prediction of 
the random phase approximation (RPA) theory \cite{Fertig},
there is no softening of the dispersion curve (roton)
at any finite momentum for any interlayer distance.
Remember that the RPA theory predicts
a roton around $k l_0 \simeq 1.4$, which was believed to 
cause an instability toward charge density wave order.
This already shows that previous theories have some serious flaws 
in describing the phase transition at $d=d_c$.
We will show in the next paragraph 
that there are completely different
low-energy excitations at large $d/l_0$.

Third, 
several previous numerical studies \cite{Moon,Chakraborty}
were unfortunately based on the exact diagonalization
of $N=8$ system, 
which, under careful investigation \cite{comment1}, 
can be shown 
to predict a Wigner crystal at large $d$
rather than a quantum Hall liquid, 
which is a finite-system artifact of boundary condition
and special number of electrons.

Fourth,
these lowest-excitation energies are exactly zero 
in the $d\rightarrow\infty$ limit
where there is no interaction between 
electrons with different pseudospins.

However, 
the most crucial aspect of the dispersion curve is that 
there are completely new low-energy excitations
due to the pseudospin spiral order, 
which are denoted by arrows in Fig.\ref{fig2}.
The lowest energy state 
in the momentum channel ${\bf k}l_0=\sqrt{2\pi/N}(N_{\downarrow},0)$
(denoted by arrows in Fig.\ref{fig2})
begins to be ripped out of the well-defined dispersion curve
at $d/l_0 \simeq 0.5$, and  it becomes completely separated from all the other
excitations at $d/l_0 \gtrsim 1.5$
so that it becomes really the lowest energy excitation 
among all momentum channels.
The same behavior is found in all studied systems 
with different values of $N$ and $N_{\uparrow}$,
with exception of the $N=8$ system 
due to a finite-size artifact as mentioned in the above.
Note that there are three other
degenerate excitations in the momentum channels
${\bf k}l_0=\sqrt{2\pi /N}(N_{\uparrow},0)$,
$\sqrt{2\pi /N}(0,N_{\downarrow})$, and
$\sqrt{2\pi /N}(0,N_{\uparrow})$
due to reflection symmetries and 
periodic boundary conditions of torus geometry.
These peculiar excitations are completely different from 
usual rotons because 
(i) they form isolated points rather than 
a part of conventional dispersion curve, and
(ii) their momentum increases as $|{\bf k}|l_0 \propto \sqrt{N}$, 
which diverges in the thermodynamic limit.
We will show later that these peculiar properties are
precisely the properties of pseudospin spiral state.

\begin{figure}
\includegraphics[width=2.3in]{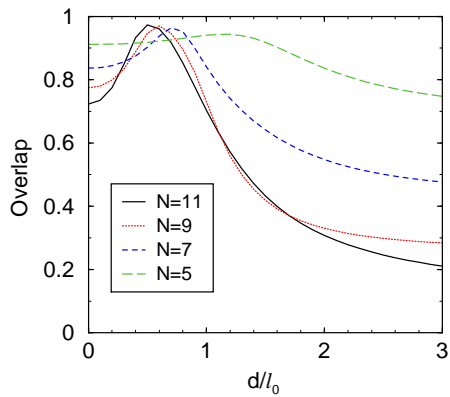}
\caption{Overlap between the pseudospin spiral wavefunction 
(described in the text)
and the lowest energy state in the Hilbert space of momentum channel
${\bf k}l_0=\sqrt{2\pi/N}(N_{\downarrow},0)$.
\label{fig3}}
\end{figure}

In order to understand why these excitations are 
the pseudospin spiral states, 
it would be best if we first examine 
the following pseudospin spiral wavefunction:
\begin{equation}
|\psi_{spiral} (n)\rangle =
\prod_m \left[ \frac{1}{\sqrt{2}}(c^{\dagger}_{m+n,\uparrow}
+c^{\dagger}_{m\downarrow}) \right] |0\rangle 
\label{spiral}
\end{equation}
where $m$ denotes the linear momentum
in the Landau gauge, i.e., $p_x=2\pi m /L$ 
and $L$ is the linear system size 
(note that $L=l_0 \sqrt{2\pi N}$ at $\nu_T=1$).
We will show later in Fig.\ref{fig3}
by explicitly computing the overlap between 
$|\psi_{spiral} (n=1)\rangle$
and the lowest-energy excitations
(denoted by arrows in Fig.\ref{fig2})
that the lowest-energy excitations are the pseudospin spiral states. 
But, first, we would like to emphasize that
the pseudospin spiral state $|\psi_{spiral} (n)\rangle$ 
is the exact ground state in the presence of parallel 
magnetic field \cite{Yang}.
A surprising result of this article is that, 
even without parallel magnetic field, 
spiral states are the lowest energy excitations 
which cause the instability of pseudospin ferromagnetic order.

\begin{figure}
\includegraphics[width=2.6in]{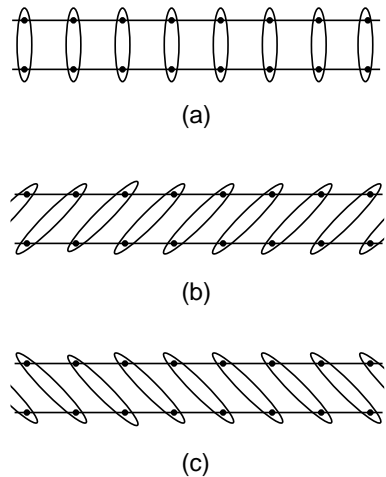}
\caption{Schematic diagram for 
(a) the pesudospin ferromagnetic state
(note the vertical interlayer correlation),
(b) the pseudospin spiral state with
clockwise spiraling direction, and
(c) the pseudospin spiral state with
anti-clockwise spiraling direction.
Ellipses indicate the Coulomb correlation of neutral pairs
between electron in one layer and hole in the other layer.
It is shown in the text that
there is an interlayer coherence due to
pseudospin spiral order at $d > d_c$,
where the ground state is conjectured to be a bound state of
two pseudospin spiral states with
opposite winding direction, 
i.e. a bound state of (b) and (c). 
\label{fig4}}
\end{figure}

To elucidate the physical meaning of pseudospin spiral state,
examine Eq.(\ref{spiral}) which reveals that 
$|\psi_{spiral}\rangle$ contains diagonal interlayer correlations
between the states with $p_x=2\pi (m+n) /L$ and $2\pi m /L$. 
Remember that the interlayer correlation
between $p_x=2\pi (m+n)/L$ and $2\pi m/L$
is identical to the diagonal interlayer correlation
between electrons with different pseudospins
which are separated laterally in the $y$-direction
by $n l_0\sqrt{2\pi/N}$.
In turn, 
this diagonal interlayer correlation is physically equivalent to
the pseudospin spiral order with spiraling period of $L/n$ because
the pseudospin part of wavefunction 
is essentially given by
\begin{equation}
\frac{1}{\sqrt{2}}
\left( 
\begin{array}{c} 
e^{i\frac{2\pi n}{L}x}  \\ 
1  
\end{array}
\right). 
\end{equation}
Pseudospin spiral excitations 
(and therefore the diagonal interlayer correlation)
are schematically depicted in Fig.\ref{fig4}.

Now we compute the momentum of pseudospin spiral states
by directly applying the magnetic translation operator 
onto $|\psi_{spiral}(n)\rangle$.
The explicit algebra shows that  
the momentum of $\psi_{spiral}(n)$ is 
$|{\bf k}| l_0=nN_{\uparrow}\sqrt{2\pi/N}$
which is exactly the momentum of 
the lowest-energy excitation 
seen in Fig.\ref{fig2} if $n=1$. 
Note that the momentum of 
this pseudospin spiral state
is actually proportional to its spiraling period.
That is the reason why the momentum of
pseudospin spiral state with long spiraling period
diverges in the thermodynamic limit.
Similarly, we can explain why the pseudospin
spiral state occurs in an isolated momentum channel.
Consider two pseudospin spiral states with long period, 
say $n=1$ and $n=2$, whose momenta
differ by a factor proportional to $\sqrt{N}/l_0$. 
Therefore, the pseudospin spiral states
with two different long spiraling periods
do not occur in adjacent momenta.

It is important to note that $|\psi_{spiral} (n=1)\rangle$
is the pseudospin spiral excitation
in the limit of long spiraling period ($=L$),
which initiates the destruction of pseudospin ferromagnetic order.
Therefore, it may be indicative of 
a second order phase transition \cite{comment2}.

As mentioned earlier, Fig.\ref{fig3} shows
the overlap between 
the pseudospin spiral state
$|\psi_{spiral} (n=1)\rangle$ in Eq.(\ref{spiral})  
and the lowest-energy excited state with 
${\bf k}l_0=\sqrt{2\pi/N}(N_{\uparrow},0)$ \cite{comment3}.
As one can see, 
the overlap is high overall at small $d/l_0$, 
and is very close to unity ($\simeq 0.97$)
especially around $d/l_0=0.5$ 
which coincides with the point 
where $\langle S_x \rangle$ begins to deviate
from the fully saturated value of $1/2$. 
This shows that
the instability of pseudospin ferromagnetic order 
is indeed caused by pseudospin spiral order.

It should be noted that,
though the overlap is lowered at larger $d/l_0$,
the low energy excitation is in general
adiabatically connected to the pseudospin spiral state.
In fact, the reason for lower overlaps at larger $d/l_0$
is that our trial wavefunction in Eq.(\ref{spiral})
is not very accurate anymore for larger $d/l_0$
since it is based on the assumption that
the ground state is the (1,1,1) state.
Of course, the ground state itself is affected by
the pseudospin spiral instability which, after all, 
destroys the pseudospin ferromagnetic order
at the critical interlayer distance.  
Therefore, it would be satisfactory 
if, in addition to the evidence due to the low energy excitation,
one can prove directly that 
the ground state has the pseudospin spiral order
at $d \simeq d_c$.
The next section will be devoted to 
the pseudospin spiral order in the ground state.

\begin{figure}
\includegraphics[width=2.6in]{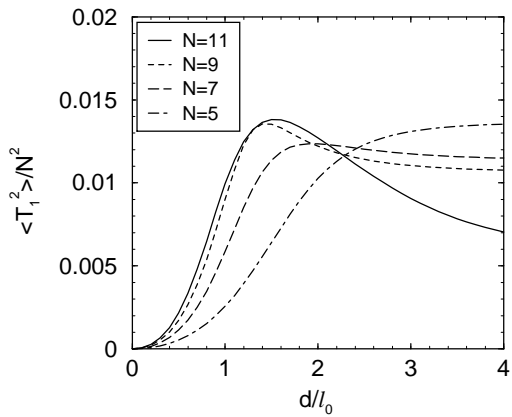}
\caption{Ground state expectation value of
the new order parameter ($T^2_n$ with $n=1$) 
measuring the diagonal interlayer correlation. 
Note that the diagonal interlayer correlation is physically equivalent to 
the pseudospin spiral order.
\label{fig5}}
\end{figure}   

\section{Pseudospin spiral order in the ground state}
\label{op}

It is conjectured that 
the ground state at large $d$
is in fact a bound state of two pseudospin spiral states
with opposite winding direction,
i.e. a bound state of (b) and (c) in Fig.\ref{fig4},
which requires a complicated interaction 
between the two spiral states
(note that this conjecture is consistent with the fact that
the ground state itself does not break translational symmetry). 
So instead of constructing a trial wavefunction
we take a more transparent approach
by defining a new order parameter 
which measures the pseudospin spiral order
(or equivalently the diagonal interlayer correlation):
\begin{equation}
T_n \equiv \frac{1}{2}
\sum_{m} \left( c^{\dagger}_{m+n,\uparrow} c_{m\downarrow}
+ c^{\dagger}_{m\downarrow} c_{m+n,\uparrow} \right)
\label{Tn}
\end{equation}

The following properties of $T_n$ are worth mentioning. 
(i) $T_n$ is a generalization of 
pseudospin magnetization $S_x$
which measures the vertical interlayer correlation: 
$T_{n=0}=S_x$.
So the relationship between $T_n$ and $\psi_{spiral}(n)$
is analogous to that between $S_x$ and $\psi_{(1,1,1)}$.

(ii) When acting upon momentum eigenstates,
$T_n$ alters momentum of the eigenstates, 
though changes in momentum are negligible 
in the thermodynamic limit.
So, in order to avoid any confusion,
we compute $\langle T^2_n \rangle$ instead.

(iii) In order to be meaningful in the thermodynamic limit,
$\langle T^2_n \rangle$ should be scaled as $N^2$.

(iv) Since the (1,1,1) state does not have 
any pseudospin spiral order, 
$\langle\psi_{(1,1,1)}|T^2_n|\psi_{(1,1,1)}\rangle/N^2$ 
should be zero in the thermodynamic limit. 
However, there are some finite-system size corrections:
$\langle \psi_{(1,1,1)} | T^2_n | \psi_{(1,1,1)} \rangle=
N/4 -N_{\uparrow}N_{\downarrow}/2(N-1) = {\cal O}(N)$.
We subtract these finite system size corrections from
the expectation value of $T^2_n$.

Fig.\ref{fig5} plots the ground state expectation value of
$T^2_{n=1}$ as a function of interlayer distance $d/l_0$.
As one can see,
$\langle T^2_{n=1} \rangle/N^2$ is peaked 
around $d/l_0 \simeq 1.5$ which is consistent with
the critical interlayer distance $d_c$.
So it is shown that indeed
the ground state contains the pseudospin spiral order
at $d \simeq d_c$.
In fact, Fig.\ref{fig5} implies that
the pseudospin ferromagnetic order and the pseudospin spiral order
may coexist for a range of $d$ since 
the expectation value of pseudospin spiral order parameter
is not zero even in interlayer distances smaller than $d_c$.
Of course, in the thermodynamic limit,
$\langle T^2_{n=1} \rangle/N^2$ 
will be zero at $d \rightarrow \infty$
since there is no interlayer correlation 
in the two split Fermi seas of composite fermions.

Finally, we would like to discuss the physical origin
of pseudospin spiral order. First, let us remind ourselves
that the pseudospin ferromagnetic state can be viewed as
a condensate of neutral pairs of electrons and correlation holes
which are bound directly across the interlayer separation.
Now for sufficiently large $d$ 
the ground state has the pseudospin spiral order,
or equivalently diagonal interlayer correlation,
which can be also viewed as a condensate of 
neutral electron-hole pairs with extended size 
and higher symmetry such as $p$-wave,
as compared to the $s$-wave-like pair in pseudospin ferromagnet.
Now, the origin of spiral order can be identified 
with the origin of extended pairs which is due to weak pairing.
Of course, the pairing between electrons and correlation holes is 
weakened as d increases. Eventually the binding is 
completely lost when the split composite fermion Fermi seas are formed.

\section{Conclusions}
\label{conclusions}

We conclude by mentioning some experimental implications.
Recently, Kellogg {\it et al.} observed
an enhanced longitudinal drag resistance
at $d/l_0$ as large as $2.6$ 
where the enhanced tunneling conductance is absent \cite{Kellogg2}. 
Theoretically,
tunneling is renormalized from 
the bare single particle tunneling gap $t$
to the greatly enhanced renormalized
tunneling $t \langle S_x \rangle$.
Therefore, it is natural to assume that
the enhanced tunneling conductance becomes absent 
as soon as the pseudospin ferromagnetic order is destroyed
at $d=d_c$.
On the other hand, we have shown that 
there is a new interlayer phase coherence 
due to the pseudospin spiral order for $d>d_c$
which, we speculate, causes the longitudinal drag anomaly
because of remaining interlayer correlation.
Regarding the Hall resistance,
our pseudospin spiral state is expected to be incompressible,
as shown in the dispersion curve.
However, the lowest energy gap is estimated to be very small
so that the Hall plateau will be difficult to be observed
in current temperature ranges and impurity concentrations.

The author is very grateful to S. Das Sarma 
for his careful reading of manuscript and 
his constant support throughout this work.
The author is also indebted to 
E. Demler, S. M. Girvin, A. Kaminski and V. W. Scarola
for their insightful comments.
This work was supported by ARDA.



\begin{thebibliography}{}


\bibitem{Spielman} I. B. Spielman, J. P. Eisenstein, 
L. N. Pfeiffer, and K. W. West, Phys. Rev. Lett. {\bf 84}, 5808 (2000).

\bibitem{Jain} J. K. Jain, Phys. Rev. Lett. {\bf 63}, 199 (1989).

\bibitem{Kalmeyer} V. Kalmeyer and S.-C. Zhang, Phys. Rev. B {\bf 46},
9889 (1992).

\bibitem{HLR} B. I. Halperin, P. A. Lee, and N. Read, Phys. Rev. B
{\bf 47},7312 (1993).

\bibitem{Scarola} V. W. Scarola, and J. K. Jain,
Phys. Rev. B {\bf 64}, 085313 (2001).

\bibitem{Fertig} H. A. Fertig, Phys. Rev. B {\bf 40}, 1087 (1989). 


\bibitem{MacDonald} A. H. MacDonald, P. M. Platzman, and G. S. Boebinger,
Phys. Rev. Lett. {\bf 65}, 775 (1990).


\bibitem{Moon} K. Moon, H. Mori, Kun Yang, S. M. Girvin,
A. H. MacDonald, L. Zheng, D. Yoshioka, Shou-cheng Zhang, 
Phys. Rev. B {\bf 51}, 5138 (1995).


\bibitem{Yang} Kun Yang, K. Moon, Lofti Belkhir, H. Mori, S. M. Girvin, 
A. H. MacDonald, L. Zheng, D. Yoshioka, 
Phys. Rev. B {\bf 54}, 11644 (1996).


\bibitem{He} S. He, S. Das Sarma, and X. C. Xie, Phys. Rev. B,
{\bf 47}, 4394 (1993).

\bibitem{Schliemann} J. Schliemann, S. M. Girvin, and A. H. MacDonald,
Phys. Rev. Lett. {\bf 86}, 1849 (2001).


\bibitem{Demler} D.-W. Wang, E. Demler, and S. Das Sarma, 
cond-mat/0303324 (2003).


\bibitem{torus} D. Yoshioka, B. I. Halperin, and P. A. Lee,
Phys. Rev. Lett. {\bf 50}, 1219 (1983); 
F. D. M. Haldane, Phys. Rev. Lett. {\bf 55}, 2095 (1985).


\bibitem{Nonlocal} As shown later, 
the ground state at large $d$ acquires
a pseudospin spiral order 
which is associated with diagonal interlayer correlation.
Since this diagonal interlayer correlation
creates a {\it nonlocal} correlation 
between laterally separated electrons (in top layer)
and correlation holes (in bottom layer)
for sufficiently large $d$,
it is no longer valid to map
the original long-range Coulomb Hamiltonian
to a usual spin model 
which is based on the existence of {\it local} spin operator.
Therefore, there is no fundamental reason for the Goldstone mode 
at sufficiently large $d$.



\bibitem{111}
Remember that the usual representation of Halperin's (1,1,1) state,
$\psi_{1,1,1}=\prod_{i<j\in\uparrow}(z_i -z_j)
\prod_{k<l\in\downarrow}(z_k -z_l)
\prod_{m\in\uparrow,n\in\downarrow }(z_m -z_n)
\prod_{p} \exp{(-|z_p|^2 /4)}$,
depicts only the orbital part of full wavefunction
which was given first in Ref.\cite{Halperin}.

\bibitem{Halperin}
B. I. Halperin, Helv. Phys. Acta {\bf 56},75 (1983).


\bibitem{Chakraborty} T. Chakraborty and P. Pietil\"{a}inen,
Phys. Rev. Lett. {\bf 59}, 2784 (1987).


\bibitem{comment1} We have proved that for $N=8$
the energy dispersion in the first Brillouin zone has
a pattern associated with rectangular-lattice Wigner crystal
which is commensurate with rectangular unit cell.
Also, this pattern persists even in the $d\rightarrow\infty$ limit. 



\bibitem{comment2} In terms of the viewpoint that
the interlayer coherent state is  
a macroscopic condensate of neutral electron-hole pairs, 
the pseusospin spiral instability with long spiral period
can be interpreted as a gradual change 
in the nature of neutral pairs
from tightly bound $s$-wave-like pairing 
to a pairing with extended size and higher symmetry.
And this continuous change in the nature of pairing
may indicate a second order phase transition
which is consistent with the viewpoint 
from the pseusospin spiral instability with long spiral period.


\bibitem{comment3} Note that 
in order to compute the overlap
one should project $|\psi_{spiral}\rangle$
into the Hilbert space of fixed $S_z$
since $|\psi_{spiral}\rangle$ is a linear combination of
various $S_z$ eigenstates similar to the (1,1,1) state.


\bibitem{Kellogg2} M. Kellogg, J. P. Eisenstein, 
L. N. Pfeiffer, and K. W. West, cond-mat/0211502 (2002).


\end{thebibliography}
\end{document}